\documentstyle[12pt]{article}
\begin{document}
\begin{flushright}
PRA-HEP-93/01\\
February 1993
\end{flushright}
\vspace{5ex}
\begin{center}
{\LARGE \bf
Strong renormalization scheme dependence in $\tau$-lepton decay:
fact or fiction?}\\[0.8in]
{\sc Ji\v{r}\'{\i} Ch\'{y}la} \\[0.1in]
{\it Institute of Physics, Academy of Sciences of Czech Republic}\\
{\it Prague, Czechoslovakia}\footnote{Postal address: Na Slovance 2,
180~40~~Prague 8, Czech Republic}\\
\vspace{.8in}
{\bf Abstract}\\
\end{center}
\begin{quotation}
The question of the renormalization scheme dependence of the
$\tau$ semileptonic decay rate is revisited in response to a
recent criticism. Particular attention is payed to a distinction
between a consistent quantitative description of this dependence and
the actual selection of a subset of ``acceptable'' renormalization
schemes. It is argued that a reasonable universal measure of the
renormalization scheme dependence can be formulated, which gives
encouraging results when applied to various physical
quantities, including the semileptonic $\tau$ decay rate.
\end{quotation}
\renewcommand\thefootnote{\arabic{footnote}}
\newpage
%\newpage \section{Introduction} %\spacing{1.5}
In a recent article
      \cite{Raczka} P. R\c{a}czka has questioned the optimism
of several earlier papers \cite{Ja,Diberder,Pich} concernig the
possibility to use the semileptonic
                                decay rate of the $\tau$-lepton and
namely the quantity $r_{\tau}$ appearing in the measured ratio
\begin{equation}
R_{\tau}\equiv \frac{\Gamma(\tau^-\rightarrow
                                   \nu_{\tau}+{\rm {hadrons}})}
       {\Gamma(\tau^-\rightarrow \nu_{\tau}{\rm e}^-\overline
{{\nu}}_{e})}=3\left(1+r_{\tau}\right)\left(1+O(\alpha_{em})\right)
\label{Rtau}
\end{equation}
for an accurate determination of the basic QCD scale parameter
$\Lambda_{\overline{\rm {MS}}}$. Let me recall that this optimism is
based on the following two favourable circumstances:
\begin{enumerate}
\item strong suppression of the potentially dangerous nonperturbative
  corrections \cite{Braaten,BNP}
\item weak renormalization scheme (RS) dependence \cite{Ja,Diberder}
\end{enumerate}

The author of \cite{Raczka} claims, on the other hand, to have found
for $r_{\tau}$
a strong RS dependence, at both the next-to-leading (NLO) and
next-to-next-to-leading (NNLO) orders. He thus comes to the
conclusion that it is impossible to use this quantity
for an accurate determination of $\Lambda_{\overline{\rm {MS}}}$.
He motivates his own analysis of the RS-dependence by the observation
that the previous analysis \cite{Ja,Diberder}, have considered
only some of the possible RS's, while
                                  ``...for a full estimate of the RS
dependence ambiguity one should compare predictions in all schemes
which {\it a priori} seem to be admissible...''.
                                              The reason for writing
this note is that although this observation is basically correct
it, if taken literally, leads to complete arbitrariness
of perturbative QCD predictions for {\it any} physical quantity.
 The $\tau$ decay rate
plays certainly no special role in this respect. I shall, moreover,
argue that the situation is not so hopeless and that one can
                                                         formulate
a plausible strategy how to construct reasonably ``accurate''
perturbative QCD predictions for many measurable quantities,
 including the $\tau$ decay rate.

The whole discussion, ongoing for more than a decade, of the RS
dependence problem
does in fact boil down to the question of finding some {\it
physically motivated} algorithm {\bf restricting} this
arbitrariness. There can be no true solution of this problem,
merely more or less reasonable assumptions can be made on
how to select one, in one way or another ``optimal'', RS
     \cite{PMS,Grunberg,BML}, and how to measure the ``strength''
of the RS dependence, i.e. how to estimate the ``theoretical'' error
associated with such a choice.

Before this can be attempted it is, however, necessary to have a
consistent quantitative description of all the degrees of freedom
associated with general renormalization group transformations.
Choosing such a consistent labelling
                                  has nothing to do with the actual
problem of ``resolving'' the RS dependence (in the above sense), but
is an inevitable prerequisite
                          for this latter step. There is no problem
in this respect and there are various distinct, though in principle
equivalent, ways how to do it. The lack of clear distinction between
these two separate issues is, however, a frequent phenomenon and can be
detected also in \cite{Raczka}.

In the notation of \cite{Raczka} (with the exception of $Q$ which is
used below to denote a generic external momentum
variable instead $m_{\tau}$, appropriate to (\ref{Rtau})),
I shall discuss QCD
predictions for any physical quantity depending on $Q$ and
admitting perturbative expansion of the form (in massless QCD)
\begin{equation}
r(Q)=a(\mu)\left(1+r_{1}(\mu/Q)a(\mu)+r_{2}(\mu/Q)a^{2}(\mu)+\cdots
\right).
\label {rta}
\end{equation}
The renormalized couplant $a(\mu)\equiv g^2(\mu)/4\pi^2$ obeys the
equation
\begin{equation}
\frac{{\rm d}a(\mu)}{{\rm d}\ln\mu}=-\beta_0a^2\left(1+c_1a(\mu)+
c_2a(\mu)^2+\cdots\right),
\label {da}
\end{equation}
where the first two coefficients $c_1,c_2$, are fixed by the number
of quark flavours, but all the higher ones
are essentially free, defining the so called {\it renormalization
convention} (RC), RC=$\{c_i,i\geq2\}$. Although not written out
explicitly, both the coefficients $r_k$ and the couplant $a(\mu)$
(when (\ref{da}) is considered to the k-th order) do, however,
depend on all $c_i,i\leq k$ as well. Finally, they also depend on the
specification which of the infinite number of solutions to (\ref{da})
we have in mind. Combining this last information with that on
$c_i,i\leq k$ defines what is called the {\it referential
renormalization scheme} (RRS) (for detailed discussion of all these
points see, for instance, \cite{my}).
Only if this RRS is fixed does the specification of the scale $\mu$
{\bf uniquely} determine the RS.
 Instead of (\ref{rta}) we can write equivalently
\begin{equation}
r(Q)=a(kQ)\left(1+r_{1}(k)a(kQ)+r_{2}(k)a^{2}(kQ)+\cdots \right),
\label{R(k)}
\end{equation}
where the arbitrariness in $\mu$ has been traded for that of the
the dimensionless parameter $k\in(0,\infty)$. As the essence of
R\c{a}czka's claim concerns both the NLO and NNLO, I shall in the
rest of this note concentrate on the NLO case and furthermore set,
purely for reasons of technical simplicity, $c_1=0$ in
(\ref{da}). All the following considerations hold for the realistic
values of $c_1$ as well, only the formulae are more cumbersome.

Under these circumstances the solutions to (\ref{da}) assume
particularly simple form
\begin{equation}
a(\mu,{\rm {RRS}})=\frac{1}{\beta_0\ln(\mu/\Lambda_{\rm {RRS}})},
\label{anlo}
\end{equation}
where the parameter $\Lambda_{\rm {RRS}}$ uniquely specifies the
solution to (\ref{da}).
In this way of labelling the RS the selection of the RRS is just a
matter of bookkeeping and so it is commonly accepted to use the
$\overline{\rm {MS}}$ RRS for this purpose. Once this convention is
made the scale $\mu$, or the parameter $k$, can be used to label
uniquelly {\bf all}
          the avaliable RS's. In view of the arbitrariness in the
selection of the RRS, no absolute meaning can, however,
be given to the scale $\mu$ or the parameter $k$ and thus no
arguments on the existence of a ``natural'' scale, given for instance by
$Q$, are sufficient to fix the RS.

I have repeated these simple facts as the author of \cite{Raczka}
starts his criticism of the ``conventional'' approach with a
correct but obvious and well-known observation just made, i.e.
that the same $\mu$,
                 or equaivalently $k$, implies different $r_i$'s and
$a(\mu)$ in different RRS's. Varying $k$ around unity, or in any other
finite interval,  in any single RRS is certainly not sufficient to
take properly into account all the RS available at the NLO, but
{\bf without} this restriction on $k$ any such
     RRS is equally suitable for a consistent labelling of all RS's.

It may, however, be preferrable to use such a labelling of RS's which
avoids
the concept of RRS and uses just one variable to uniquely fix a RS.
One way of doing this is to use for this purpose the value of $r_1$
itself. The internal consistency of perturbation theory implies the
following relation between $\mu, a$ and $r_1$:
\begin{equation}
r_1=b\ln\left(\frac{\mu}{\Lambda_{\overline{\rm {MS}}}}\right)-\rho=
\frac{1}{a}-\rho\;\;\Rightarrow a=\frac{1}{r_1+\rho}
\label{a(r)}
\end{equation}
where $\rho$ is the renormalization group invariant \cite{PMS}.
  It is solely through this invariant that the
               $Q$-dependence of $r(Q)$ enters (\ref
{rta}). Substituting (\ref{a(r)}) into (\ref{rta}) and truncating
it to the NLO we get
\begin{equation}
r^{\rm {NLO}}(\rho,r_1)=\frac{2r_1+\rho}{(r_1+\rho)^2}
\label{rnlo}
\end{equation}
as an explicit function of $r_1$ and $\rho$. In Fig.1 the dependence
of $r^{\rm {NLO}}$ on $r_1$
is displayed for several values of $\rho$ (only those points for
which (\ref {rnlo}) stays positive are plotted).
%%%%%%%%%%%%%%%%%%% figure 1 starts here
%\begin{figure}
%\begin{center}
%\fbox{\epsfig{file=raczka1.eps,height=10cm}}
%\caption{
%$r^{\rm {NLO}}_{\tau}(r_1,\rho)$ as a function of $r_1$ for several
%values of $\rho$.}
%\end{center}
%\end{figure}
%%%%%%%%%%%%%%%%%% end of Fig.1
So far only the question of an exhaustive and consistent labelling
of the RS's has been discussed. Now the crucial moment arrives and
and some point (or some interval of points) on the curve
corresponding to  a given $\rho$ must be selected. Looking on the
curves in Fig.1, we conclude that:
\begin{description}
\item
i/ there is one ``exceptional'' point on each of the curves, namely
   the stationary point, defing the PMS \cite{PMS} choice of the RS
   (in our approximation this point coincides with the effective
   charges approch (ECH) of \cite{Grunberg}, defined by the condition
   $r_1=0$).
\item
ii/The pattern of $r_1$ dependence is qualitatively the same for
   all values of $\rho$.
\end{description}
To give the word ``strong'' or ``weak'' RS dependence a
quantitative content requires that we define the range of
of ``acceptable'' $r_1$.
 In \cite{Ja} we have taken as our preferred RS
 the PMS/ECH one and furthermore suggested to estimate the associated
``theoretical error'' by the difference
\begin{equation}
\Delta^{theory}=r^{\rm {NLO}}({\rm {PMS}})
                              -r^{\rm {NLO}}(\overline{\rm {MS}}).
\label{error}
\end{equation}
This is of course somewhat arbitrary  definition
and we could certainly take some other measure of the RS-dependence.
We, however, consider it meaningful to assume this, or some other,
definition of the theoretical ``error''
and use it in analyses of all physical quantities for which the
NLO and, if possible, also the NNLO, calculations are available.
For $\Lambda_{\overline {\rm {MS}}}$ in the region of a few
hundreds of MeV and taking into account that, for 3 quark flavours,
$r_1(\mu=Q,\overline {\rm {MS}})=5.2$ the range of $\rho$
appropriate
to the $\tau$ decay rate (\ref{rta}) is roughly between 2 and 4.

After the $\tau$-lepton decay rate and ${\rm e}^+{\rm e}^-$
annihilation into hadrons
\cite{Ja} we have recently analyzed in the same way the
Gross-Llewellyn-Smith sum rule \cite{my} and intend to continue in
                                                        this
direction as further NNLO calculations become available. I stress the
importance of comparing this kind of analysis at NLO and NNLO as
only this comparison can tell us whether our choice of the RS leads to
a reasonable convergence (in the pragmatical sense) as we proceed to
higher orders and whether the associated ``theoretical'' error
simultaneously decreases as it should if our procedure is sound.
The results of \cite{Ja,my} are encouraging in both respects.

  The dependence of  $r^{\rm {NLO}}$ on $r_1$ and $\rho$
as given in (\ref{rnlo}) is quite general and holds for any physical
quantity admitting perturbative expansion of the form (\ref{rta}).
The difference between various physical quantities enters entirely
through the corresponding values of $\rho$ due to the fact that
\begin{equation}
\rho=b\ln\left(\frac{Q}{\Lambda_{\overline{\rm {MS}}}}\right)
                                     -r_1(\mu=Q,
 \overline{\rm {MS}})
\label{rho}
\end{equation}
contains both possible differences in the scale $Q$ and
                                in $r_1(\mu=Q,\overline{\rm MS})$.

%%%%%%%%%%%%%%%%% Fig.2 starts here
%\begin{figure}
%\begin{center}
%\fbox{\epsfig{file=raczka2.eps,height=12cm}}
%\caption{
%$r^{\rm {NLO}}_{\tau}(r_1,\rho)$ as a function of $\rho$ for several
%values of $r_1$ in the region appropriate to (a) the $\tau$ decay rate
%  and (b) the R-ratio in e$^{+}$e$^{-}$ annihilations.}
%\end{center}
%\end{figure}
%%%%%%%%%%%%%%%%%% end of Fig.2
In Fig.2a the formula (\ref{rnlo}) is plotted as a function of $\rho$
for a number of values of $r_1$. It can be redrawn (as in
\cite{Raczka}) for any particular physical quantity expressible in
the form (\ref{rta}) as a function of $b\ln\left(Q/
\Lambda_{\overline{\rm {MS}}}\right)$ by
simply shifting the origin of Fig.2a by an appropriate value of $r_1$
and taking into account different value of external momentum $Q$.
The content of Fig.2a, with curves corresponding to
$r_1\in(-3.83,8.32)$
has been interpreted in \cite{Raczka} as an evidence for the
strong RS dependence of the $\tau$ decay rate (\ref{rta}).
                This interval of $r_1$ values correspond to the
overlap between those obtained within the $\overline{\rm {MS}}$
RRS varying $k$ in (\ref{R(k)}) between 1/2 and 2 and between 1/3 and
1 within the symmetric MOM RRS.
No argument has, however, been put forward to justify the restriction
to the above mentioned RRS and thus the selected interval of $r_1$
values must be considered as arbitrary as any other one.
It must be born in mind that
all values of $r_1$ are in principle equally acceptable and that even
restricting the values of $k$ in (\ref{R(k)}) to the region around
unity we can always find such a RRS in which $r_1$ is equal to any
prescribed value. Consequently
also the conclusions drawn in \cite{Raczka} and based on this
restriction of ``acceptable'' RS is valid only within this particular
definition. Taking into account sufficiently large upper and
lower bounds on $r_1$ would, of course, make any physical quantity
``strongly'' RS-dependent.

To illustrate this point let me consider the same expression (\ref
{rnlo}) in the region $\rho \in(10,20)$, which is roughly the range
covered by the PETRA experiments measuring the familiar R-ratio
\begin{equation}
R_{{\rm e}^+{\rm e}^-}\equiv \frac{\Gamma({\rm e}^+
                                             {\rm e}^-\rightarrow
{\rm {hadrons}})}{\Gamma({\rm e}^+{\rm e}^-\rightarrow \mu^+\mu^-)}
= 3(1+r_{{\rm e}^+{\rm e}^-})
\label{epem}
\end{equation}
The quantity $r_{{\rm e}^+{\rm e}^-}$ differs from (\ref{rta}),
beside the range of $\rho$ only
by the value of $r_1(\mu=Q,\overline{\rm {MS}})=1.41$. In Fig.2b
    the same curves as in Fig.2a are plotted
in this high $\rho$ region ( I
continue to labell as ``$\overline{\rm {MS}}$''
                              the curve corresponding
to $r_1=5.2$ despite the fact that for (\ref{epem})
      $r_1(\mu=Q,\overline{\rm {MS}})=1.41$ in
order to maintain direct relation to Fig.2a). While the spread of
the results corresponding to $r_1\in(-3.83,8.32)$ has
decreased with respect to Fig.2a, extending this
interval of ``acceptable'' $r_1$ just a little bit down to
moderately negative values around -5 or -6, would imply ``strong''
RS dependence even for the R-ratio (\ref{epem})! I don't think there
is any argument why $r_1=-3.83$ should be acceptable while
$r_1=-5$ not. And even at LEP energies (corresponding to $\rho$
around 22) $r_1=-11$ would be sufficient to
                           yield $r_{{\rm {e}}^+{\rm {e}}^-}=0$.

                        In the preceding paragraphs
I have argued in favour of a particular measure of the
strength of the RS-dependence. Compared to the measure based on
 an ad hoc choice of the
interval of $r_1$, it has another advantage, which appears
when going beyond the NLO. There, new parameters, one at each
further order, are needed to describe the full RS dependence.
We can choose $r_i; i\geq2$, (or $c_i; i\geq2$) for that purpose
but in any case if we attempt to define the ``acceptable'' RS by
means of restrictions on these further parameters, new criteria
have to be invented at each new higher order. For instance in \cite
{Raczka} $c_2$ was considered in the interval $c_2\in(-25,25)$, with
no particular reason given for these limits. In our approach,
based on the selection of PMS/ECH and $\overline{\rm {MS}}$ RS's,
we don't face such problems as these criteria are reasonably
(although not entirely) free of ambiguities.

Summarizing this note, it is fair to say that the considerations
presented above are, or at least should be, nothing new.
But neither is there any new idea in \cite{Raczka}.
There is no real solution to the RS
dependence ambiguity,
                  short of calculating the expansions to all orders.
I have tried to stress the distinction between the task of a full
and consistent description of this ambiguity, which presents no
problem and the actual task of selecting one, or a subset,
of the RS's. In this latter
                     step some arbitrariness and subjective choice
is inevitable. Bearing this in mind the conclusions of \cite{Raczka}
have their validity only within a particularly defined measure of the
strength of the RS-dependence adopted by its author. They should
certainly not discourage further attempts, both theoretical and
experimental, to use the $\tau$ decay rate (\ref{Rtau}) as a
potentially suitable place for quantitative tests of perturbative QCD.

\vspace{1cm}
%\newpage

%\end{document}
\vspace {0.9cm}
%\newpage
\parindent 0.cm
{{\large \bf Figure captions.}}\\

\vspace {0.7cm}
 Fig.1:
$r^{\rm {NLO}}_{\tau}(r_1,\rho)$ as a function of $r_1$ for several
values of $\rho$.

\vspace{0.6cm}
Fig.2:
$r^{\rm {NLO}}_{\tau}(r_1,\rho)$ as a function of $\rho$ for several
values of $r_1$ in the region appropriate to  \\
\parindent 1.cm
(a) the $\tau$ decay rate and (b) the R-ration in e$^+$e$-$
  annihilations.


\begin{thebibliography}{99}
\bibitem {Raczka} P. R\c{a}czka, Phys. Rev. D {\bf 23}, R3699 (1992)
\bibitem {Ja} J. Ch\'{y}la, A. Kataev, and S. Larin, Phys. Lett. B
   {\bf 267}, 269 (1991)
\bibitem {Diberder} F. Le Diberder and A. Pich, Phys. Lett. B {\bf 289}
  , 165 (1992)
\bibitem {Pich} A. Pich, in {\it Proceedings of the XXVIIth
  Rancontre de Moriond}, Les Arcs, March 1992, edited by Tran Than Van
 (Editions Frontieres, Gif-sur-Yvette, 1992)
\bibitem {Braaten} E. Braaten, Phys. Rev. Lett. {\bf 60}, 1606 (1988);
 {\bf 63}, 577 (1989)
\bibitem {BNP}
E. Braaten, S. Narison, and A. Pich, Nucl. Phys. B {\bf 373},
 581 (1992)
\bibitem {PMS} P. M. Stevenson, Phys. Rev. D {\bf 23}, 2916 (1981)
\bibitem {Grunberg} G. Grunberg, Phys. Rev. D {\bf 29}, 2315 (1984)
\bibitem {BML} S. Brodsky, G. P. Lepage, and P. Mackenzie,
   Phys. Rev. D {\bf 28}, 228 (1983)
\bibitem {my} J. Ch\'{y}la and A. Kataev, Phys. Lett. B {\bf 297}, 385
                            1992
\end{thebibliography}
\end{document}